\documentclass{Interspeech2024}

\usepackage{hyperref}
\usepackage{cleveref}
\usepackage{caption}
\usepackage{multirow}
\usepackage{algorithm}
\usepackage{algorithmic}
\usepackage{setspace}

\interspeechcameraready

\title{A Transcription Prompt-based Efficient Audio Large Language Model for Robust Speech Recognition}

\name[affiliation={1,2}\dagger]{Yangze}{Li}
\name[affiliation={2}\dagger]{Xiong}{Wang}
\name[affiliation={2}]{Songjun}{Cao}
\name[affiliation={2}]{Yike}{Zhang}
\name[affiliation={2}]{Long}{Ma}
\name[affiliation={1}*]{Lei}{Xie}

\address{
  $^1$Audio, Speech and Language Processing Group (ASLP@NPU), \\ Northwestern Polytechnical University, China \\
  $^2$Tencent YouTu Lab, China}
  
\email{yzli@mail.nwpu.edu.cn, chnxwang@tencent.com, lxie@nwpu.edu.cn\thanks{$\dagger$ Equal contribution, * Corresponding author.}}


\begin{document}

\maketitle
% the abstract here must exactly match the abstract entered into the paper submission system

\begin{abstract}
 Audio-LLM introduces audio modality into a large language model (LLM) to enable a powerful LLM to recognize, understand, and generate audio. 
% As an important task in audio-LLM,  ASR is a classified task that has unique paired input and label, contradictory to the generative task of LLM. 
% So the hallucination caused by LLM results in semantic correction but textual fault for the ASR task in audio-LLM and the decoding repetition of LLM will affect the performance of the ASR task seriously. 
However, during speech recognition in noisy environments, we observed the presence of illusions and repetition issues in audio-LLM, leading to substitution and insertion errors.
This paper proposes a transcription prompt-based audio-LLM by introducing an ASR expert as a transcription tokenizer and a hybrid Autoregressive (AR) Non-autoregressive (NAR) decoding approach to solve the above problems. 
Experiments on 10k-hour WenetSpeech Mandarin corpus show that our approach decreases 12.2\% and 9.6\% CER relatively on Test\_Net and Test\_Meeting evaluation sets compared with baseline. 
Notably, we reduce the decoding repetition rate on the evaluation set to zero, showing that the decoding repetition problem has been solved fundamentally.
\end{abstract}

\section{Introduction}
\vspace{4pt}
LLMs~\cite{DBLP:journals/corr/abs-2303-08774, DBLP:journals/corr/abs-2302-13971, DBLP:journals/jmlr/ChowdheryNDBMRBCSGSSTMRBTSPRDHPBAI23, DBLP:journals/corr/abs-2309-16609} based on decoder-only Transformer~\cite{DBLP:conf/nips/VaswaniSPUJGKP17} have revolutionized the field of natural language processing (NLP). 
Due to their ability to capture complex linguistic patterns and contextual information, LLMs perform impressive results on NLP tasks like machine translation, sentiment analysis, text generation, etc. 
% LLMs not only significantly enhance neural network modeling capabilities but also enable fluent human-computer text interaction, effectively reducing the gap between research and applications. 
Against this background, a significant amount of recent research has aimed at creating a seamless integration of text and audio through a unified large-scale audio-language model, enabling models to handle various tasks within and between these modalities.

Although unified audio models~\cite{DBLP:conf/emnlp/ZhangLZZWZQ23, DBLP:journals/corr/abs-2306-12925, DBLP:conf/asru/WangHSWCCCZSRZYPSSW23, DBLP:journals/corr/abs-2310-13289, DBLP:journals/corr/abs-2311-07919, DBLP:journals/corr/abs-2310-04673, DBLP:journals/corr/abs-2303-17580, DBLP:journals/corr/abs-2305-16107} have shown considerable potential in tasks such as speech translation and speech understanding, their performance in speech recognition tasks still lacks robustness compared to well-tuned expert models, particularly in speech with complex acoustic environments. 
In this paper, we have observed the following issues introduced by the LLM-based framework have led to a degradation in the performance of audio-LLM speech recognition.
% The first problem is the rich knowledge and associative ability of LLM will lead to semantic correction of recognition results but textual fault. 
The first issue is that the rich knowledge and associative abilities of LLM can lead to semantic corrections of recognition results, but may introduce substitution errors at the same time.
The second one is the audio-LLM will lead to text fragment repetition during AR decoding in speech recognition tasks.
This leads to many insertion errors and makes the recognition results difficult to comprehend.
% While the above problems can not be easily solved by some commonly used strategies such as temperature and top-p~\cite{DBLP:conf/iclr/HoltzmanBDFC20}/top-k~\cite{DBLP:conf/acl/LewisDF18} in NLP tasks because of the unique paired between transcription and speech.
Although the above issues can be addressed in NLP tasks using common strategies such as temperature and  top-p~\cite{DBLP:conf/iclr/HoltzmanBDFC20}/top-k~\cite{DBLP:conf/acl/LewisDF18}, there is currently no effective solution for these problems in speech recognition due to the need for accurate transcriptions.

\vspace{-6pt}
The main reason for the above issues is that prior works tend to introduce speech modality only through pre-trained ASR encoders but ignore information about the textual modality of speech, and the hallucination of LLM is not alleviated by a specific design for the speech recognition task. 
In this paper, inspired by such work like GER~\cite{DBLP:conf/asru/YangGLGBS23, DBLP:conf/nips/0075HYSCS23, hu2024large} which demonstrates that utilizing LLM for post-processing ASR transcriptions can also enhance recognition performance, we propose a transcription prompt-based audio-LLM that combines information from both the speech and the text modality obtained from ASR well-tuned expert models to enhance the speech recognition performance of the audio-LLM model. 
% The contribution of this paper is summarized below:
Our specific approaches to addressing these issues are as follows:
\vspace{-6pt}
\begin{enumerate}
\item[i)] 
% We propose an effective training framework for information from both modalities. 
We propose an effective training framework that utilizes both modalities.
Following the structure of Whipser~\cite{DBLP:conf/icml/RadfordKXBMS23}, we concatenate the recognition transcriptions generated by an ASR expert model as textual prompts before the speech embedding and employ special token sequences to guide the task. Our approach enhances speech recognition performance effectively by helping LLM extract semantic information from speech modalities and improving their contextual modeling capabilities through the additional transcription prompts generated from a NAR ASR expert model trained with CTC (Connectionist Temporal Classification)~\cite{DBLP:conf/icml/GravesFGS06} loss. This ASR expert model can constrain LLM from the textual modality to avoid the additional transcribe errors caused by its excessive generation ability.
\item[ii)] Since the CTC loss function establishes a time alignment between speech and text resulting in the problem of decoding repetition, we further propose a hybrid AR NAR decoding approach that uses textual prompts during the decoding step. This approach can solve the decoding repetition problem of audio-LLM fundamentally and achieves a lower ASR decoding real-time factor (RTF) by the hybrid approach.
\end{enumerate}
\vspace{-6pt}

Our proposed approach is mainly evaluated on 10k-hour WenetSpeech~\cite{DBLP:conf/icassp/ZhangLGSYXXBCZW22} Mandarin corpus. From the results, our model achieves superior performance on the Test\_Net and Test\_Meeting evaluation dataset, decreasing 12.2\% and 9.6\% on CER relatively compared with the baseline model, while significantly accelerating the decoding step with 32\% relative time reduction. 
Furthermore, our results on AISHELL-1~\cite{DBLP:conf/ococosda/BuDNWZ17} corpus indicate that our approach has strong generalization capabilities for low-cost domain adaptation. By analyzing the decoding repetition rates on each evaluation set, we reduce this rate to zero, showing that we completely solved this problem with our proposed hybrid AR NAR decoding approach.

\section{Related Work}
\textbf{Integration methods:} Decoder-only LLMs can control and influence the generated text through prompt design, hence several multimodal models based on LLMs have been developed to expand the application of LLMs beyond text-based tasks. Models like speechGPT~\cite{DBLP:conf/emnlp/ZhangLZZWZQ23} and AudioPaLM~\cite{DBLP:journals/corr/abs-2306-12925} use discrete representations such as speech tokens to help LLMs finish speech-related tasks~\cite{DBLP:journals/corr/abs-2305-16107, DBLP:journals/corr/abs-2311-04534}, but these models may lose some speech-related information due to the conversion of continuous speech signals into discrete tokens. Experiments in LauraGPT~\cite{DBLP:journals/corr/abs-2310-04673} have shown that this conversion process causes a decline in performance on speech-related tasks compared to models that use continuous speech features. Besides, fine-tuning LLMs is difficult due to the large number of parameters for LLMs hence avoid doing this in most cases. The SLM~\cite{DBLP:conf/asru/WangHSWCCCZSRZYPSSW23} and Qwen-Audio~\cite{DBLP:journals/corr/abs-2311-07919} have shown that they achieve impressive performance on various speech-related tasks while keeping the LLM frozen even though the LLMs have been fine-tuned in LauraGPT~\cite{DBLP:journals/corr/abs-2310-04673} and Salmonn~\cite{DBLP:journals/corr/abs-2310-13289}.

% \vspace{-4pt}
\noindent\textbf{Repetition Problem:} Text generation tasks in NLP usually use likelihood as a training objective to yield high-performance models. However, for decode-only models such as LLM, output text may become dull, incoherent, or stuck in repetitive loops while using maximization-based decoding approaches like greedy search. To avoid this problem, the current widely-used approach is to perform sampling strategies on the predicted probabilities during decoding, such as nucleus sampling~\cite{DBLP:conf/iclr/HoltzmanBDFC20} and top-k sampling~\cite{DBLP:conf/acl/LewisDF18}. However, for classification tasks like speech recognition, the decoding result from the model should be unique. Therefore, these probabilistic sampling strategies are not suitable for the ASR task because they may introduce additional errors, which are also verified by some experiments in this paper. In addition, the repetition problem is usually difficult to solve through some simple rules such as maximum decoding token limit, so our proposed approach gives valid guidance for solving the repetition problem in speech-related classification tasks of audio-LLM. 

\section{Proposed Method}

\subsection{Model architecture}
\begin{figure}[!t]
\centering
\includegraphics[width=0.43\textwidth]{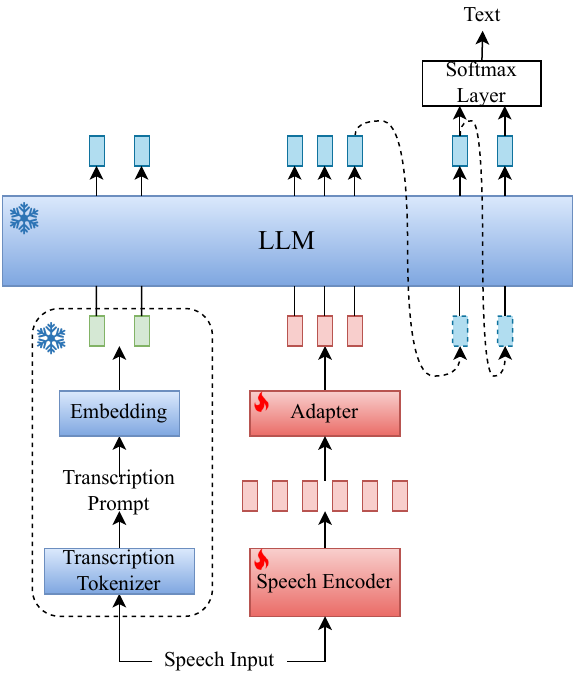}
\caption{
 	 The overview of our audio-LLM architecture. 
 	}
  \label{fig1}
\vspace{-16pt}
\end{figure}

As shown in Fig.~\ref{fig1}, the audio-LLM in our approach consists of four main components: an LLM, a transcription tokenizer, a speech encoder, and an adapter.

\textbf{LLM} Audio-LLM is built upon an LLM, which serves as its fundamental component. Based on the powerful and flexible decoder-only structure of LLM, we can easily concatenate the input sequences of text and speech modalities, allowing LLM to learn the information between the two modalities autonomously.

\textbf{Transcription tokenizer} To further discover the powerful language capabilities of LLM, we introduce a tokenizer to provide the transcription prompt for input speech. In this paper, the tokenizer is an ASR pre-trained model using CTC loss.
% consisting of 12 layers of Conformer~\cite{DBLP:conf/interspeech/GulatiQCPZYHWZW20} layers with a hidden size of 512. Besides, the CTC tokenizer takes in the 80-dimensional mel-filterbank feature with a 10 ms window shift and a 25 ms frame length. 
This tokenizer will decode input speech to text using CTC greedy search, then the text will be converted to discrete semantic representation by the text embedding layer of the LLM. 

\textbf{Speech encoder} We employ the speech encoder with the same architecture and initialization as the transcription tokenizer without the project layer of output, and this component will be trained during the training stage. Ultimately, the encoder converts roughly every segment of the original audio signal into a high-dimensional representation.

\textbf{Adapter} The adapter component connects the representations generated by the speech encoder with the text embeddings of the LLM, initializing with random weights during the training stage. This module contains such layers of 1D convolution and a fully connected layer, which maps the high-dimensional speech encoder output to the LLM text embedding. As a result, the adapter is optimized to map each segment of the input speech into the continuous semantic space of LLM.

% \textbf{Prompt format} To expand multilingual and multitasks ability, we employ a special sequence with 4 tokens to guide the task information. Especially, the format of the prompt for LLM can be described as follows: \text{“}[SOT] [ZH] [TRANSCRIBE] [ZH] [AUDIO] Audio [/AUDIO]\text{”}. When the CTC tokenizer is given, the output text of the CTC tokenizer will be added as a prefix for the prompt: \text{“}[PREFIX] transcript [/PREFIX] [SOT] [ZH] [TRANSCRIBE][ZH] [AUDIO] audio [/AUDIO]\text{”}. In which, the first [ZH] represents the source language of the speech, while the second [ZH] represents the target language of the output text.

\vspace{-2pt}
\subsection{Training framework}
\vspace{-4pt}
For speech recognition task of audio-LLM, training stage need paired data denoted as $(\mathbf{x}, \mathbf{y})$, where $\mathbf{x}$ represents speech input and $\mathbf{y}$ represents the corresponding text sequences $\left\{y_{0}, y_{1}, \cdots, y_{N-1}\right\}$. As shown in Eq.(\ref{eq1}) and Eq.(\ref{eq2}), the main objective during training is to maximize the probability of the next text token $y_{n}$ given last token sequence $\mathbf{y}_{<n}$ and high-dimensional representation $\mathbf{H}_s$ generate by speech encoder and adapter, where $\mathcal{L}_{\text{CE}}$ is the loss function to optimize.
\begin{align}
    & \mathbf{H}_s=\operatorname{Adapter}(\operatorname{Encoder}(\mathbf{x}))\label{eq1}
    \\
    & \mathcal{L}_{\text{CE}}=-\sum_{n=0}^{N-1}\log\mathcal{P}_{LLM}\left(y_{n} \mid \mathbf{y}_{<n}, \mathbf{H}_s;\Theta_{\text{LLM}}\right)   \label{eq2}
\end{align}
By taking the transcription generated by tokenizer into the prompt,  the loss function $\mathcal{L}_{\text{CE\_prompt}}$ is described as shown in Eq.(\ref{eq3}), where transcription prompt $\mathbf{y}_{\text{prompt}}=\operatorname{Tokenizer}\left(\mathbf{x}\right)$.
\begin{equation}
    \mathcal{L}_{\text{CE\_prompt}}=-\sum_{n=0}^{N-1}\log\mathcal{P}_{LLM}\left(y_{n} \mid \mathbf{y}_{<n}, \mathbf{y}_{\text{prompt}}, \mathbf{H}_s;\Theta_{\text{LLM}}\right) \label{eq3}
\end{equation}
During the training stage, to avoid the model's over-fitting on the transcription prompt, we use a hyper-parameter $\lambda\in[0, 1]$ to control whether the current utterance has a transcription prompt or not. As a result, the training loss function for each utterance in a training batch is shown in Eq.(\ref{eq4}),  
\begin{equation}
    \mathcal{L}=\left\{\begin{matrix}
    \mathcal{L}_{\text{CE\_prompt}}&&,p\leqslant \lambda\\
    \mathcal{L}_{\text{CE}} && ,p>\lambda 
    \end{matrix}\right. \label{eq4}
\end{equation}
while $p$ is a random number generated for each utterance in every training batch with a uniform distribution in [0, 1].

\begin{figure}[!h]
\vspace{-16pt}
\centering
\includegraphics[width=0.46\textwidth]{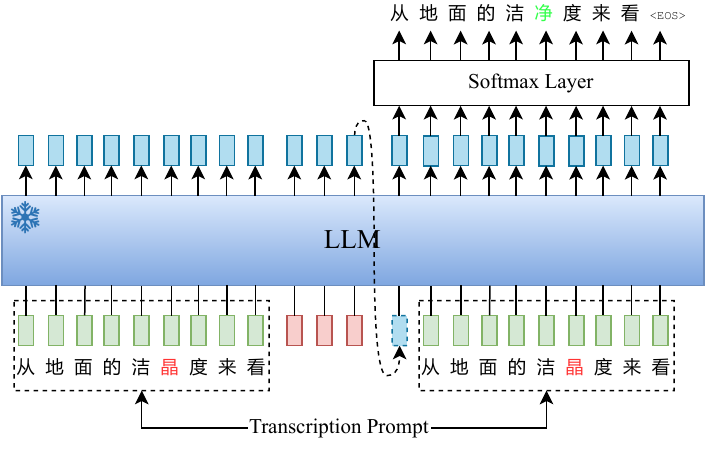}
\caption{
 	 NAR decoding approach combined with transcription prompt.
 	}
  \label{fig2}
  \vspace{-20pt}
\end{figure}

\subsection{Decoding}

\textbf{AR decoding} Audio-LLM usually uses the AR decoding approach, which predicts the next token by the last predicted token until $<\text{EOS}>$ is predicted as shown in Eq.(\ref{eq5}). In this paper, the AR decoding approach is used as default if the experiment details do not mention the type of decoding approach.
\begin{equation}
    y_{n}^{\ast }={\underset{y_{n}}{\text{argmax}}}\text{ }\mathbf{LLM}(\mathbf{y}_{<n}, \mathbf{H}_s;\Theta_{\text{LLM}}) \label{eq5}
\end{equation}

\textbf{NAR decoding}
By introducing the transcription prompt decoded by the tokenizer, we propose a fast NAR audio-LLM decoding approach.  As shown in Figure \ref{fig2}, we replace the context $\mathbf{y}_{<n}$ predicted by the LLM with ${y}_{\text{prompt}<n}$ predicted by the tokenizer. Finally, we can perform a NAR decoding approach that generates predicted text sequence $\mathbf{y}^{\ast }$ in one step described in Eq.(\ref{eq6}). 
\begin{equation}
    \mathbf{y}^{\ast }={\underset{\mathbf{y}}{\text{argmax}}}\text{ }\mathbf{LLM}( \mathbf{y}_{\text{prompt}}, \mathbf{H}_s;\Theta_{\text{LLM}}) \label{eq6}
\end{equation}

In the NAR decoding approach, it can be described that the LLM modifies the transcription prompt and plays as an error correction model. Since the length of the predicted text sequence only relies on the length of the transcription prompt, this approach will avoid the problem of repetition.

\textbf{Hybrid AR NAR decoding}
Although the NAR decoding approach can solve the repetition problem, the ability of the LLM is limited by the fixed length of the transcription prompt. To combine the advantages of AR and NAR decoding approaches, we propose a hybrid AR NAR decoding approach as shown in the algorithm~\ref{alg:1}. This hybrid approach determines whether there is a problem such as repetition in the AR decoding approach using decode length limit hyper-parameter$\sigma$, and then uses the NAR decoding result if the condition is triggered, to take into account the advantages of both AR and NAR decoding approaches. In this paper, the hyper-parameter $\sigma$ is empirically set to $1.5$.

\begin{algorithm}
    \setstretch{1}
    \caption{Pipeline of Hybrid AR NAR decoding approach}
    \label{alg:1}
    \begin{algorithmic}[1]
        \STATE {Given a well-trained proposed audio-LLM}
        \STATE {Given input speech feature $\mathbf{x}$ and decode parameter $\sigma$}
        \STATE {Compute high-dimensional representation $\mathbf{H}_s$ from Eq.(\ref{eq1}})
        \STATE {Compute transcription prompt $\mathbf{y}_{\text{prompt}}=\operatorname{Tokenizer}\left(\mathbf{x}\right)$, and get token number of $\mathbf{y}_{\text{prompt}}$ as $L_{\text{prompt}}$}
        \STATE {Initialize the decode result $\textbf{y}^{\ast}$ with an empty sequence, set length of decode result $L_{\text{decode}}$ as $0$}
        \WHILE {$\textbf{y}^{\ast}$ is not end with $<\text{EOS}>$}
        \STATE {Generate next token $y_{n}^{\ast }$ from Eq.(\ref{eq5}})
        \STATE {$L_{\text{decode}}=L_{\text{decode}}+1$}
        \IF {$L_{\text{decode}}>{\sigma}{\times}L_{\text{prompt}}$}
        \STATE {Replace $\textbf{y}^{\ast}$ with Eq.(\ref{eq6}})
        \RETURN {$\textbf{y}^{\ast}$}
        \ENDIF
        \STATE {Append $y_{n}^{\ast }$ to the end of $\textbf{y}^{\ast}$}
        \ENDWHILE
        \RETURN {$\textbf{y}^{\ast}$}
    \end{algorithmic}
    \vspace{-2pt}
\end{algorithm}
\vspace{-6pt}

\section{Experiments}
\subsection{Dataset}
\vspace{-2pt}
In this paper, we evaluate our proposed approach to the WenetSpeech corpus~\cite{DBLP:conf/icassp/ZhangLGSYXXBCZW22}. This corpus contains over 10,000 hours of high-quality labeled Mandarin speech, which is sourced from YouTube and podcasts, covering different speaking styles, scenarios, domains, topics, and noise environments. We use two carefully checked evaluation sets Test\_Net and Test\_Meeting, the first one is a match set compared with training data, and the second one is a mismatch set that contains far-field and conversational meeting speech. In addition, we use the well-known public set AISHELL-1~\cite{DBLP:conf/ococosda/BuDNWZ17} to confirm the out-of-domain performance of the model.

\vspace{-2pt}
\subsection{Experimental setup}
\vspace{-2pt}
In this paper, the LLM is initialized with pre-trained weights obtained from Qwen-7B~\cite{DBLP:journals/corr/abs-2309-16609}. Qwen-7B is a Transformer decoder model with 32 layers and a hidden size of 4096, comprising a total of 7.7 billion parameters.
We implement a two-stage training approach for our proposed transcription prompt-based LLM. For the first stage, we trained a transcription tokenizer which used a learning rate of 1e-3 with a batch size of 256, performed 5000 steps of warm-up, and employed gradient accumulation with a factor of 16. The transcription tokenizer consists of 12 layers of Conformer~\cite{DBLP:conf/interspeech/GulatiQCPZYHWZW20} layers with a hidden size of 512. Besides, the transcription tokenizer takes in the 80-dimensional mel-filterbank feature with a 10 ms window shift and a 25 ms frame length. 
For the second stage, we initialize the speech encode with the same parameters of the transcription tokenizer trained in the first stage, then we used a learning rate of 1e-4 with a batch size of 64, performed a warm-up for 2000 steps, and employed gradient accumulation with a factor of 16. Besides, for the second stage, only the speech encoder and adapter are trained while the transcription tokenizer and LLM are kept frozen. This adapter contains 2 1D-convolution layers and 1 fully connected layer, which maps the dimension of the speech encoder from 512 to 4096. In addition, the first convolution layer uses a stride of two to do down-sampling. The trainable parameters of the speech encoder amount to 70 million, of which the adapter is 10 million. When using AR greedy decoding, we set the maximum decoding token limit to 200 (as the evaluation set contained sentences with a maximum of 180 tokens).

\begin{table}[!h]
\vspace{-5pt}
\caption{CER (\%) of various models on Test\_Net, Test\_Meeting and Test\_aishell1. The RTF is computed as the ratio of the total inference time to the total duration of evaluation sets.}
\vspace{-5pt}
\centering
\scalebox{0.78}{
\begin{tabular}{lccccc}
\toprule
\hline
\multicolumn{1}{c}{\multirow{2}{*}{Model}} & \multicolumn{3}{c}{CER (\%)}                                                                      & \multicolumn{1}{l}{\multirow{2}{*}{RTF}} \\
\multicolumn{1}{c}{} & \multicolumn{1}{l}{Test\_Net} & \multicolumn{1}{l}{Test\_Meeting} & \multicolumn{1}{l}{Test\_aishell1} & \multicolumn{1}{l}{}                     \\
\hline
\textit{Baselines}        &    &   &  \\
Conformer-W1                 & 8.60  & 14.30 & 4.61 &  \\
% Whisper                   & 8.50  & 14.59 &  \\
Qwen-Audio                & 9.62  & 9.05 & 1.59 &  \\
\hline
\textit{Ours}             &  & & & \\
Audio-LLM                 &  & & & \\
% \quad +Top-3 sample      & 11.14 & 17.64 & 0.32 \\
\quad + $\lambda=0.0$    & 9.18 & 15.30 & 4.11 & \\
\quad + $\lambda=0.5$    & 8.47 & 13.94 & 3.86 & 0.39   \\
\quad \quad +NAR          & 8.35 & 14.62 & 3.83 & \textbf{0.04} \\
\quad \quad +Hybrid AR NAR  & \textbf{8.09}  & 13.83 & \textbf{3.71} & 0.25 \\
\quad + $\lambda=1.0$ & & & & \\
\quad \quad +Hybrid AR NAR & 8.26 & \textbf{13.65} & 3.73 & \\
\hline
\bottomrule
\end{tabular}}
\label{main_results}
\vspace{-12pt}
\end{table}

\begin{table}[!h]
\vspace{-5pt}
\caption{Comparison of insertion, deletion, and substitution errors among different approaches on Test\_Net.}
\vspace{-5pt}
\centering
\scalebox{0.95}{
\begin{tabular}{lccc}
\toprule
\hline
\textit{Model} & Insertion  &  Deletion  & Substitution   \\
\hline
Audio-LLM         &  &  &  \\
\ \  + $\lambda=0.0$      & 4160 & 16408 & 25503 \\
\ \  + $\lambda=0.5$      & 3732 & 12058 & 19291 \\
\ \ \ \   + NAR             & 1897 & 12035 & 20662 \\
\ \  \ \   + Hybrid AR NAR         & 2251 & 12258 & 19004 \\
\hline
\bottomrule
\end{tabular}}
\label{details}
\vspace{-15pt}
\end{table}

\subsection{Analysis on transcription prompt}
To analyze the effect of the transcription prompt for audio-LLM, we first set two baseline models Conformer-W1 and Qwen-Audio. Conformer-W1 is trained on WenetSpeech in the same setup as the CTC-based transcription tokenizer mentioned in Sec.4.2 and Qwen-Audio is an audio-LLM that has achieved a state-of-the-art speech recognition performance recently. The result of our proposed transcription prompt-based audio-LLM is shown in Tab.~\ref{main_results}. From the result, for $\lambda=0$ which means do not use transcription prompt during the training stage, the audio-LLM and Qwen-Audio get a worse result on Test\_Net due to the more insertion errors caused by the hallucination of LLM. After we introduced the transcription prompt with $\lambda=0.5$, the model got a significant improvement on evaluation sets compared with the model on $\lambda=0$. Furthermore, we also compared the effect of different decoding approaches and resulting in our proposed hybrid AR NAR decoding approach will bring additional improvement for audio-LLM during the decoding stage, even if the CER is significantly lower than Qwen-Audio on Test\_Net (9.62 $\to$ 8.09) and Conformer-W1 on Test\_aishell1 (4.61 $\to$ 3.71). In addition, we designed an ablation experiment with $\lambda=1.0$ to prove that over-reliance on the transcription prompt can take disadvantages to the audio-LLM.

To further analyze the detailed effects of the transcription prompt, we list the insertion, deletion, and substitution errors in Tab.~\ref{details}. It shows that the model with $\lambda=0.5$ has less error on the three types than the model with $\lambda=0.0$, and after using the NAR decoding approach, insertion errors decrease a lot. This shows the transcription prompt can restrain its over-generation ability for the speech recognition task. Furthermore, the proposed hybrid AR NAR decoding approach will further decrease substitution errors which shows CTC transcription prompt can improve the modal alignment ability of audio-LLM. 
% It is worth mentioning that the AR and hybrid AR NAR decoding approaches have lower RTF because they reduce the inference times of the LLM part.
It is worth mentioning that the hybrid AR NAR decoding approach allows for earlier truncation of AR decoding when repetition problems arise, resulting in lower RTF compared to the AR decoding approach.

\vspace{-5pt}
\subsection{Analysis on repetition problem}
We defined sentence-level decoding repetition ratio (DRR) as the number of sentences that fall into the repetition problem divided by the total number of the evaluation set, to measure the seriousness of the repetition problem. As shown in Tab.~\ref{repetition}, after the introduction of the transcription prompt and the hybrid AR NAR decoding approach, the DRR will be reduced to 0 step by step, which means the repetition problem is completely solved. Compared with the existing approaches, we show the result of the top-3 samples strategy. As a result, the decoding problem repetition seems to be effectively alleviated, but resulting in an unacceptable CER increase, mainly because the ASR task is a classification task rather than a generative task.

\vspace{-5pt}
\begin{table}[!h]
\caption{The sentence-level decoding repetition ratio (DRR (\textperthousand)) for each model. }
\vspace{-5pt}
\centering
\begin{tabular}{lcccc}
\toprule
\hline
\multirow{2}{*}{\textit{Model}} & \multicolumn{2}{c}{Test\_Net} & \multicolumn{2}{c}{Test\_Meeting} \\
& DRR  & CER & DRR & CER        \\
\hline
Qwen-Audio     & 0.43 & 9.62              & 0.83 & 9.05                  \\
\hline
Audio-LLM      \\
\quad + $\lambda=0.0$ & 0.36 & 9.18              & 0.96 & 15.30                  \\
\quad \quad + top-3 sample & 0.03 & 11.14       & 0.60 &17.64              \\
\quad + $\lambda=0.5$   & 0.28 &8.47              & 0 & 13.94       \\
\quad \quad + AR NAR      & 0 & 8.09              & 0 & 13.83             \\
\hline
\bottomrule
\end{tabular}
\label{repetition}
\vspace{-18pt}
\end{table}

\begin{table}[!h]
\caption{CER (\%) of different models on Test\_Net and AISHELL-1, when $\lambda=0.5$ for audio-LLM and hybrid AR NAR decoding approach is used. }
\vspace{-5pt}
\centering
\scalebox{0.9}{
\begin{tabular}{lccc}
\toprule
\hline
\textit{Model}   & Tokenizer & Test\_Net & Test\_aishell1 \\
\hline
Conformer-W1        & -              & 8.60      & 4.61     \\
% W1               & -              & 8.63      & -        \\
Conformer-A1              & -              & 50.13         & 5.20     \\
Audio-LLM-W1 & Conformer-W1              & 8.09      & 3.71     \\
Audio-LLM-W1 & Conformer-A1             & 12.56         & \textbf{3.08}     \\
% \hline
% \multirow{3}{*}{\begin{tabular}[c]{@{}l@{}} \quad + transcription prompt \end{tabular}} 
%         & Conformer      & 8.09      & 3.86     \\
%         % & W1             & \textbf{7.96}      & -        \\
%         & A1             & -         & \textbf{3.08}     \\
% \hline
\bottomrule
\end{tabular}}
\label{generation}
\end{table}
\vspace{-15pt}

\subsection{Generalization for robustness}
To further evaluate how the transcription prompt affects the audio-LLM, we provided transcription prompts generated by another tokenizer different from the one during training. As shown in Tab.~\ref{generation}, audio-LLM-W1 means the model uses WenetSpeech to train and initialize the tokenizer and speech encoder, and Conformer-A1 means the ASR expert model trained with AISHELL-1 corpus. The results show different tokenizers will lead the audio-LLM to related domains, proving our proposed approach has the robustness to achieve domain adaptation. 

\vspace{-5pt}
\section{Conclusion}
In this paper, we proposed a transcription prompt-based audio-LLM for the ASR task. Specifically, we introduced a transcription tokenizer to generate transcription prompts for audio-LLM and a hybrid AR NAR decoding approach to avoid the hallucination and repetition problems of audio-LLM for the ASR task. As a result, our proposed approach evaluated on WenetSpeech can decrease 12.2\% and 9.6\% on CER relatively on the Test\_Net and Test\_Meeting compared with the baseline model, and we reduce the sentence-level decoding repetition ratio to zero, resulting we completely solved the repetition problem. Besides, our generalization validation experiment also shows our proposed method has the ability of low-cost domain adaptation for audio-LLM.

% \section{Acknowledgements}

\bibliographystyle{IEEEtran}
\bibliography{mybib}

\end{document}